# Evolution of magnetic properties in the vicinity of the Verwey transition in $Fe_3O_4$ thin films

X. H. Liu,[1,2,*] W. Liu,[2] and Z. D. Zhang[2]

[1] *Max Planck Institute for Chemical Physics of Solids, Nöthnitzerstraße 40, 01187 Dresden, Germany*

[2] *Shenyang National Laboratory for Materials Science, Institute of Metal Research, Chinese Academy of Sciences, Shenyang 110016, China*

We have systematically studied the evolution of magnetic properties, especially the coercivity and the remanence ratio in the vicinity of the Verwey transition temperature ($T_V$), of high-quality epitaxial Fe3O4 thin films grown on MgO (001), MgAl2O4 (MAO) (001), and SrTiO3 (STO) (001) substrates. We observed rapid change of magnetization, coercivity, and remanence ratio at $T_V$, which are consistent with the behaviors of resistivity versus temperature [$\rho(T)$] curves for the different thin films. In particular, we found quite different magnetic behaviors for the thin films on MgO from those on MAO and STO, in which the domain size and the strain state play very important roles. The coercivity is mainly determined by the domain size but the demagnetization process is mainly dependent on the strain state. Furthermore, we observed a reversal of remanence ratio at $T_V$ with thickness for the thin films grown on MgO: from a rapid enhancement for 40-nm- to a sharp drop for 200-nm-thick film, and the critical thickness is about 80 nm. Finally, we found an obvious hysteretic loop of coercivity (or remanence ratio) with temperature around $T_V$, corresponding to the hysteretic loop of the $\rho(T)$ curve, in Fe3O4 thin film grown on MgO.

## I. INTRODUCTION

Magnetite (Fe3O4) is one of the most important transition-metal oxides and is being actively investigated due to its rather unique and interesting set of electrical and magnetic properties, such as the high Curie temperature (858 K), relatively high saturation magnetization, small coercivity field [1,2], theoretically predicted half-metallic behavior [3–6], and the famous first-order metal-insulator transition at 124 K (known as the Verwey transition) [7]. These make Fe3O4 a potential candidate for spin electronic devices, such as spin valves or spin tunnel junctions [8–14]. However, it has been very difficult to realize these unique properties in thin films for various device applications, owing to the uncontrollable formation of growth defects, such as antiphase boundaries (APBs) [15,16]. It has been reported that the APBs generally exist in epitaxial Fe3O4 thin films grown on MgO, sapphire, MgAl2O4, or SrTiO3 substrates due to the lower symmetry and twice the lattice constant of Fe3O4 (Fd3m) with respect to MgO (Fm3m) and the large lattice mismatch for the other substrates [14–50]. Numerous studies have shown that because of the presence of APBs in $Fe_3O_4$ thin films, their transport and magnetic properties strongly deviate from the single-crystal bulk, such as the low Verwey transition temperature ($T_V$) and very broadened transition, and an increased resistivity with decreasing film thickness [14,18–27,29–35,37,38], negative unsaturated magnetoresistance [14,18–23,25–27,33], superparamagnetic behavior in ultrathin films [41,42], and non-saturation magnetization at very high fields [15–17,46,47]. Recently, Liu *et al.* [51] obtained exceptionally high quality $Fe_3O_4$ thin films grown on tailored spinel $Co_{2-x-y}Mn_xFe_yTiO_4$ substrates with relatively small lattice mismatch, which not only exhibit the Verwey transition as sharply as the single-crystal bulk but also present quite high $T_V$ up to 136.5 K. This work gives an example of the better Verwey transition in $Fe_3O_4$ thin films than in the single-crystal bulk. It is clear that the $Fe_3O_4$ thin films without APBs show intrinsic transport properties, which are quite different from those of films grown on general substrates [14,18–27,29–35,37,38]. However, the magnetic $Co_{2-x-y}Mn_xFe_yTiO_4$ substrates greatly restrict the study in magnetic properties of these $Fe_3O_4$ thin films.

The magnetic properties of bulk $Fe_3O_4$ have been widely studied in the past several decades [52–60]. With lowering temperature across the Verwey transition, the transformation from cubic to monoclinic structure leads to a sharp jump for some magnetic parameters at $T_V$, such as the rapid drop of magnetization and the sharp enhancement of coercivity. On the other hand, although a tremendous amount of work has been devoted to investigating the magnetic properties of $Fe_3O_4$ thin films [14–18,20,33,34,39–42,44–50], the systematic study of magnetic properties of the thin films in the vicinity of the Verwey transition has been seldom reported [44,45]. Bollero *et al.* [44] investigated the influence of thickness on microstructural and magnetic properties in $Fe_3O_4$ thin films. They gave out the variation of coercivity and remanence across the $T_V$, and the APBs are considered to significantly affect the magnetization behaviors of $Fe_3O_4$ thin films. In previous work, however, the Verwey transition in $Fe_3O_4$ thin films was very broad and the $T_V$ was quite low; the quality of these thin films should not be high enough (the presence of chemical off stoichiometry or many APBs), which would greatly influence the magnetic properties. In our previous work, we have grown quite high quality epitaxial $Fe_3O_4$ thin films; the Verwey transition is very sharp and the $T_V$ reaches to 122 K for 200-nm-thick $Fe_3O_4$ film on MgO

(001) [37], which is close to that of the single-crystal bulk and much higher than in previous work [15–17,41–50]. Therefore, to more comprehensively understand the magnetic behaviors in Fe₃O₄ thin films, it is very necessary to reinvestigate the magnetic properties in high-quality Fe₃O₄ thin films.

In this work, we systematically investigate the evolution of magnetic properties with temperature of high-quality epitaxial Fe₃O₄ thin films grown on MgO (001), MgAl₂O₄ (MAO) (001), and SrTiO₃ (STO) (001) substrates. Rapid changes of magnetization, coercivity, and remanence ratio were observed at $T_V$, which are in agreement with the behaviors of resistivity versus temperature [$\rho(T)$] curves for different thin films. Especially, the thin film grown on MgO was found to exhibit quite different magnetic behavior from that on MAO and STO, in which the domain size and the strain state take the dominate role. The coercivity is mainly determined by the domain size but the demagnetization process is mainly related to the strain state. The small domain size and high strain state lead to large coercivity and incoherent reversal of magnetization, respectively. Furthermore, a very interesting reversal of remanence ratio with thickness, from a rapid increase for 40-nm- to a sharp drop for 200-nm-thick film, at $T_V$ was noticed for the thin films grown on MgO, and the critical thickness is about 80 nm, which might be greatly related to interface effect of the thin films. At last, a clear hysteretic loop of coercivity (or remanence ratio) with temperature around the $T_V$, corresponding to the loop of the $\rho(T)$ curve, was observed in Fe₃O₄ thin film grown on MgO.

## II. EXPERIMENTAL

Fe₃O₄ thin films were grown by molecular beam epitaxy (MBE) in an ultrahigh vacuum (UHV) chamber with a base pressure in the $1 \times 10^{-10}$ mbar range. High-purity Fe metal was evaporated from a LUXEL Radak effusion cell at temperatures of about 1250 °C in a pure oxygen atmosphere onto single-crystalline MgO (001), MgAl₂O₄ (001) (MAO), and SrTiO₃ (001) (STO) substrates. These substrates were annealed for two hours at 600 °C in an oxygen pressure of $3 \times 10^{-7}$ mbar to obtain a clean and well-ordered surface structure before the Fe₃O₄ deposition. Standard samples were grown using an iron flux of 1 °A/min, an oxygen background pressure of $1 \times 10^{-6}$ mbar, and a growth temperature of 250 °C [37]. *In situ* and real-time monitoring of the epitaxial growth was performed by reflection high-energy electron diffraction (RHEED) measurements. Oscillations in the RHEED specular beam intensity, where each oscillation corresponds to the formation of one new atomic monolayer (ML), allows for precise control of the film thickness. The crystalline structure was also verified *in situ* after the growth by low-energy electron diffraction (LEED). The RHEED patterns were taken at 20 keV electron energy, with the beam aligned parallel to the [100] direction of the substrate and the LEED patterns were recorded at an electron energy of 88 eV. Furthermore, all samples were analyzed *in situ* by x-ray photoelectron spectroscopy (XPS). The XPS data were collected using 1486.6 eV photons (monochromatized Al $K\alpha$ light) in normal emission geometry and at room temperature using a ScientaR3000 electron energy analyzer. The overall energy resolution was set to about 0.3 eV. Measurements for temperature dependence of transport and magnetic properties of $Fe_3O_4$ thin films were performed with the standard four-probe technique using a physical property measurement system (PPMS) and a superconducting quantum interference device (SQUID), respectively. X-ray diffraction (XRD) was employed for further *ex situ* investigation of the structural quality and the microstructure of the samples. The XRD measurements were performed with a high-resolution PANalytical X'Pert MRD diffractometer using monochromatic Cu $K\alpha_1$ radiation ($\lambda = 1.54056$ A°).

### III. MICROSTRUCTURAL CHARACTERIZATION

Figure 1 shows the RHEED patterns of clean substrates MgO (001) [Fig. 1(a)], MAO (001) [Fig. 1(b)], and STO (001) [Fig. 1(c)], and RHEED and LEED patterns of 200-nm-thick $Fe_3O_4$ thin films grown on MgO [Figs. 1(d) and 1(g)], MAO [Figs. 1(e) and 1(h)], and STO [Figs. 1(f) and 1(i)], respectively. The sharp RHEED streaks and the presence of Kikuchi lines, as well as the intense and sharp LEED spots [Figs. 1(d) and 1(g)] indicate a flat and well ordered (001) crystalline surface structure of the $Fe_3O_4$ thin film on MgO. Because the growth is fully coherent, with the in-plane dimensions of the spinel unit cell of $Fe_3O_4$ being twice those of the rock-salt unit cell of MgO, one expects a new set of diffraction rods (spots) occurring with half spacing of the substrate. The RHEED and LEED patterns indeed show the occurrence of the half-order diffraction rods (spots) in the zeroth Laue zone. The signature of the ($\sqrt{2} \times \sqrt{2}$)$R45°$ surface reconstruction, which is also characteristic for single-crystal magnetite, corresponding to a new set of diffraction rods and spots, is clearly observed in Figs. 1(d) and 1(g), respectively. At the same time, the typical RHEED and LEED images with the presence of ($\sqrt{2} \times \sqrt{2}$)$R45°$ surface reconstruction are obviously seen for the films on MAO [Figs. 1(e) and 1(h)] and on STO [Figs. 1(f) and 1(i)], demonstrating also the high quality of these two $Fe_3O_4$ thin films. Moreover, we found that for the film grown on STO, the RHEED and LEED patterns become slightly weak and blurred, indicating a slightly increasing disorder in this film due to the large lattice mismatch (−7.5%). We have recorded the time development of the crystalline structure during the films growth. In Fig. 2 we can clearly observe pronounced intensity oscillations, which are indicative of a two-dimensional layer-by-layer growth; the time period of the oscillation is 58 s, which corresponds to the time needed to grow 1 ML of $Fe_3O_4$ and allows for a precise thickness determination.

In order to clarify the chemical state of the iron oxide, the wide scan with binding energy from 1200 to−18 eV and Fe $2p$ core level XPS spectra were *in situ* recorded for 200-nm-thick $Fe_3O_4$ thin films grown on MgO (001), MAO (001), and STO (001), as shown in Figs. 3(a) and 3(b), respectively. The XPS spectra in Figs. 3(a) and 3(b) demonstrate a quite clean surface and represent the typical signatures of $Fe_3O_4$ thin films [37,51,61,62]. The identical XPS spectra also indicates no influence of substrates on the chemical composition of the thin films. XRD experiments have been *ex situ* performed on 200-nm-thick $Fe_3O_4$ thin films grown on MgO, MAO, and STO, respectively, as shown in Fig. 4. It is found that the XRD patterns do not display any phase other than substrates and $Fe_3O_4$. For the samples grown on MgO (red color) and on STO (olive color), the (002)∕(004) and (004)∕(008) reflections correspond to MgO (or STO)/$Fe_3O_4$ because of the lattice constant of $Fe_3O_4$ as twice as that of MgO (or STO), and the (004) and (008) reflections for both MAO and $Fe_3O_4$ are presented for the blue color pattern. Moreover, with respect to the samples on MAO and STO, the $Fe_3O_4$ peaks of the sample on MgO obviously shift to a high degree due to the fully tensile strained state with reduction of lattice constant **c**. The fully relaxed state for samples on MAO and STO because of large lattice mismatch results in the same lattice constant, corresponding to the same degree $\theta$, as that of the bulk; the strain state of the samples will be shown later. XRD rocking-curve measurements of the in-plane reflection (115) of the $Fe_3O_4$ thin films on MgO, MAO, and STO are displayed as an inset of Fig. 4. A sharp peak with quite small full width at half maximum (FWHM) can be seen for $Fe_3O_4$ thin film on MgO, while the samples grown on MAO and STO exhibit very broad peaks. By using the simple Scherrer formula [63], we calculated an average domain size of 98 nm, 20 nm, and 11 nm for 200-nm-thick $Fe_3O_4$ thin films grown on MgO, MAO, and STO, respectively, which is similar to our previous work [37].

Furthermore, the strain state of the Fe$_3$O$_4$ thin films was also studied by XRD measurements. Figure 5 presents the reciprocal space mapping of the (226) and (115) reflections of 200-nm-thick Fe$_3$O$_4$ films grown on MgO (001) (a) and MAO (001) (b), respectively. Any lattice relaxation will result in a deviation of the (226) or (115) line; i.e., the satellite peaks will not line up with the substrate peak located at (226) and (115). Here, a relaxation rate can be defined as $R = (\mathbf{a}_{film} - \mathbf{a}_{substrate})/(\mathbf{a}_{bulk} - \mathbf{a}_{substrate})$, where $\mathbf{a}_{film}$ and $\mathbf{a}_{substrate}$ are the in-plane lattice parameters of the Fe$_3$O$_4$ thin film and substrate, respectively, and $\mathbf{a}_{bulk}$ is the lattice constant for bulk Fe$_3$O$_4$ (8.397 Å). According to this definition, $R = 0\%$ and $100\%$ will correspond respectively to a fully strained and fully relaxed state of the thin film. In Fig. 5, fully strained and fully relaxed states can be obtained for 200-nm-thick Fe$_3$O$_4$ thin films grown on MgO [Fig. 5(a)] and MAO [Fig. 5(b)], respectively. Therefore, the critical thickness is bigger than 200 nm for Fe$_3$O$_4$ thin film grown on MgO, which is much larger than that calculated from the Fischer, Kuhne, and Richter model [64], but similar to that reported by Arora *et al.* [28]. The 200-nm-thick Fe$_3$O$_4$ thin film grown on MAO is fully relaxed due to the much larger lattice mismatch (−3.7%) [32,64]. However, in Bollero *et al.*'s work [44], the 320-nm-thick Fe$_3$O$_4$ film grown on MAO (001) still shows an 80% relaxation rate.

## IV. TRANSPORT CHARACTERIZATION

Having investigated the microstructural characterization of the 200-nm-thick Fe$_3$O$_4$ thin films grown on different substrates, we then measured the transport properties of these samples. The resistivity as a function of temperature $\rho(T)$ for the 200-nm-thick Fe$_3$O$_4$ films grown on MgO (001), MAO (001), and STO (001) are presented in Fig. 6. Here the Verwey transition temperatures $T_{V+}$ and $T_{V-}$ are defined as the temperature of the maximum slope of the log[$\rho(T)$] curve for the warming up and cooling down temperature branches, respectively. It is observed that the film grown on MgO displays very sharp transition with $T_{V+}$ of 122 K, close to that of the single-crystal bulk, whereas the transition becomes very broad with $T_{V+}$ of around 124.5 K for the film grown on MAO or STO. From the XRD results we have obtained an average domain size of about 98 nm, 20 nm, and 11 nm for 200-nm-thick Fe$_3$O$_4$ films grown on MgO, MAO, and STO, respectively (see inset of Fig. 4). The observation of a broad transition implies more inhomogeneity for the latter two samples; i.e., they consist of a wide distribution of crystallites each having its own transition temperature [37]. The broadest Verwey transition induced from the smallest average domain size is seen for the film on STO, and a big hysteresis of about 4 K for the film on MgO would be attributed to the large density of APBs [37]. To better understand the transport properties of the Fe$_3$O$_4$ thin films, we fitted the $\rho(T)$ curves above the Verwey transition (from 130 to 300 K) with an Arrhenius law: $\rho(T) = \rho_\infty \exp(E_a/kT)$ [21,30] and nonadiabatic polaron model: $\rho(T) = \rho_0 T^{3/2} \exp(E_a/kT - C/T^3 + D/T^5)$ [65]; here, $E_a$ is an activation energy, and $C$ and $D$ are constant parameters. It is found that the fitting results are much better by using the latter model (see the inset of Fig. 6), and the activation energy is about 32 meV (about 43 meV by the Arrhenius law) for 200-nm-thick Fe$_3$O$_4$ film grown on MgO, which is much lower than 56 meV in Ref. [21] and 46 meV in Ref. [30]. At the same time, the Fe$_3$O$_4$ film grown on MAO or STO has a higher activation energy of 46 meV or 71 meV due to its much smaller average domain size. Therefore, the smaller the average domain size, the larger the activation energy, and the more difficult the hopping of the electrons in Fe$_3$O$_4$ thin films above $T_V$.

## V. MAGNETIC CHARACTERIZATION

As shown in the above sections, the microstructural characterization and the transport property measurements indicate the high quality of our Fe$_3$O$_4$ thin films. We began the investigation of magnetic properties of Fe$_3$O$_4$ thin films grown on different substrates. Figure 7 shows zero-field and field cooling (ZFC-FC) magnetization as a function of temperature for 200-nm-thick Fe$_3$O$_4$ thin films grown on MgO (001) (a), MAO (001) (b), and STO (001) (c), respectively, in an applied field of 200 Oe. Rapid change of magnetization at $T_{V+}$ can be clearly seen for each ZFC curve, whereas the behavior of the FC curve is quite different for the films on MgO and on MAO or STO. An obvious reduction of magnetization at $T_{V-}$ can be also observed for the film on MgO [Fig. 7(a)] but the variation of magnetization is very broad for the films on MAO or STO. Especially, one notices in Figs. 7(b) and 7(c) that the magnetization first increases to a maximum peak and then slowly drops with lowering temperature across $T_{V-}$. The $dM/dT$ as a function of temperature near $T_V$ [see insets of Figs. 7(a), 7(b) and 7(c)] shows two peaks at $T_{V+}$ and $T_{V-}$, which are nearly consistent with those from $\rho(T)$ curves in Fig. 6, while only a clear peak at $T_{V-}$ can be seen for the film on MgO. Therefore, the behaviors of magnetization with temperature are greatly in agreement with the transport properties of the films, in which the domain size takes a very important role.

After studying the temperature dependence of magnetization for different thin films, we then focused on the change of magnetic properties as a function of magnetic field at different temperatures. In this process, the system was first zero-field-cooled from 300 to 30 K and then the magnetic hysteresis loops $M(H)$ at different constant temperatures were measured with temperature warming up from 30 to 300 K. The in-plane ($H//$ along the [100] direction) $M(H)$ curves at different temperatures for 200-nm-thick $Fe_3O_4$ film on MgO are shown in Fig. 8(a); obviously different magnetic properties can be seen at temperatures below and above $T_{V+}$. At $T$ <120 K or $T$ >123 K, the variation of coercivity ($H_C$) or remanence ($M_r$) is very small; when it reaches the vicinity of $T_{V+}$, the $H_C$ and the magnetization rapidly changes (see 121, 122, and 123 K). The $dM/dH$ vs $H$ shown as an inset of Fig. 8(a) presents a peak at $H_C$ for each curve; the sharpest peak of about 3 emu/$cm^3$ $Oe^{-1}$ with the smallest full width at half maximum (FWHM) of 130 Oe can be seen for 123 K ($>T_{V+}$), while the broadest peak of only 1 emu/$cm^3$ $Oe^{-1}$ with largest FWHM of 310 Oe is found for 122 K ($=T_{V+}$). It is clear that the reversal of domains with applied magnetic field is slowest and steps across the largest field range at Verwey transition, and the relatively gradual reversal of magnetization for 121 K ($<T_{V+}$) is due to the much larger monoclinic magnetocrystalline anisotropy constant [60]. At $T>T_{V+}$, the $H_C$ becomes quite small but the low-field magnetization is much larger than that at $T<T_{V+}$.

For comparison, the in-plane magnetic behaviors of 200-nm-thick $Fe_3O_4$ film grown on MAO shown in Fig. 8(b) are quite different. The $M(H)$ curves display almost rectangular shape at different temperatures including $T_{V+}$ in which the domains are coherently reversed at $H_C$. Relative to the results in Fig. 8(a), with increasing temperature, we can observe a big decrease of $H_C$ from 120K but small change of magnetization. The low-field magnetization and the $M_r$ slightly reduce at $T_{V+}$, which is much smaller than that of the film on MgO [see Fig. 8(a)]. Interestingly, $H_C$ reaches a minimum at 130 K (120

Oe) and then gradually enhances with further increasing temperature up to 240 Oe at 300 K. At the same time, the in-plane magnetic behavior of the 200-nm-thick Fe$_3$O$_4$ film on STO shown in Fig. 8(c) is similar to that on MAO [Fig. 8(b)]. The nearly rectangular shape of the $M(H)$ curves can be observed also at different temperatures though the average domain size of this sample is only 11 nm. The $Hc$ gradually decreases with increasing temperature and gets to a minimum value also at 130 K, and then increases to 330 Oe at 300 K. The variation trend of the low-field magnetization and the $Mr$ around the Verwey transition is nearly the same as that in Fig. 8(b).

To further understand the magnetic properties in Fe$_3$O$_4$ thin films, we investigated the evolution of out-of-plane ($H_\perp$ along [001] direction) magnetic properties with temperature for Fe$_3$O$_4$ thin films grown on MgO (001), MAO (001), and STO (001). The Fe$_3$O$_4$ films grown on these three substrates have very good perpendicular anisotropic properties at different temperatures. Similarly, the out-of-plane results also show an obvious variation of some magnetic parameters at $T_V$. As the out-of-plane $Mr$ is very small, the change of these values is not very sharp, especially for the films on MAO and STO, around $T_V$. The normalized magnetic hysteresis loops for in-plane and out-of-plane cases in an applied field of 50 kOe at 50 and 300 K, respectively, of 200-nm-thick Fe$_3$O$_4$ film on MgO is shown in Fig. 8(d). Relative to the in-plane case, the out-of-plane $M(H)$ curves show typical hard axis magnetic behaviors at 50 and 300 K respectively: large $Hc$, quite small $Mr$, and low-field magnetization, which can be clearly seen in the inset of Fig. 8(d). With respect to a sharp peak of $d(M/M_S)/dH$ at $Hc$ for in-plane, even no peak can be found for the out-of-plane case, meaning a quite incoherent change of magnetic moments with applied magnetic field (hard axis). To better present the variation of magnetic properties around the Verwey transition, we plotted the in-plane and out-of-plane magnetic hysteresis loops at 120 K (below $T_V$) and 130 K (above $T_V$) for 200-nm-thick Fe$_3$O$_4$ films grown on MgO (001) [Figs. 9(a) and 9(d)], MAO (001) [Figs. 9(b) and 9(e)], and STO (001) [Figs. 9(c) and 9(f)], respectively. Clearly, the easy (hard) axis always stays in-plane (out-of-plane) for the three samples.

The in-plane and out-of-plane coercivity ($Hc$) and remanence ratio ($Mr/Ms$) as a function of temperature for 200-nm-thick Fe$_3$O$_4$ films grown on MgO (001), MAO (001), and STO (001) are summarized in Figs. 10(a) and 10(d), 10(b) and 10(e), and 10(c) and 10(f), respectively. We first discuss the in-plane condition; as can be seen the film on MgO has the sharpest jump of $H_c$ at $T_V$, the value rapidly changes from 410 Oe at 121 K to 80 Oe at 123 K, and then it slightly reduces to only 50 Oe at 300 K, while $H_c$ remains nearly constant at 450 Oe at $T$ <120 K. However, the change of $H_c$ around $T_V$ becomes very broad for the film on MAO or STO [see Figs. 10(b) and 10(c)]; it varies from 520 Oe or 450 Oe at 120K to 120 Oe or 200 Oe at 130 K, and then increases to 190 Oe or 330 Oe at 200K but remains nearly constant from 200 to 300 K, respectively. Moreover, one notices that the values of $H_c$ in Figs. 10(b) and 10(c) are much larger than those in Fig. 10(a) at different temperatures, and remarkably the minimum of $H_c$ occurs at 130 K for the former two samples. On the other hand, the out-of-plane $H_c$ at different temperatures is larger than that of in-plane for each sample due to the hard axis. At the same time, the evolution of $M_r/M_s$ with temperature also presents a rapid change at $T_V$ [see Figs. 10(d), 10(e) and 10(f)]. Furthermore, the sharpest variation of $M_r/M_s$ can be seen for the film on MgO both at in-plane and out-of-plane cases. As the good perpendicular anisotropic properties for all the films, the values of $M_r/M_s$ for out-of-plane at different temperatures are quite small (only around 0.1). In Fig. 10(d), the in-plane $M_r/M_s$ is around 0.58 at low temperatures, which jumps to a minimum of 0.35 at 122 K and then gradually increases to around 0.5 with further rising temperature. However, the values of in-plane $M_r/M_s$ for the film on MAO are much larger, remaining nearly constant at about 0.78 at $T$ <120 K and reducing to about 0.7 at $T_V$, and then becoming nearly constant above $T_V$ [Fig. 10(e)]. Similarly, the film on STO in Fig. 10(f) shows also very large in-plane $M_r/M_s$ of about 0.7 below $T_V$ and about 0.65 above $T_V$, respectively. In Fig. 10 it is found that the evolution of $H_c$ and $M_r/M_s$ in the vicinity of $T_V$ is consistent with the electrical behaviors of the $\rho(T)$ curves in Fig. 6; the magnetic behavior for the film on MgO is quite different from that on MAO and STO.

The rapid change of $H_c$ with temperature across $T_V$ can be easily understood. As with the abrupt enhancement in magnetocrystalline and magnetostriction constants when $Fe_3O_4$ transforms from cubic ($Fd\bar{3}m$) to monoclinic ($Cc$) structure, the value of the dominating monoclinic magnetocrystalline anisotropy constant is about 10 times greater than the cubic anisotropy constant $K_1$, and thus a sharp increase of $H_c$ is noticed at $T_V$ [see Figs. 10(a), 10(b) and 10(c)]; the broad variation of $H_c$ corresponds to the broad Verwey transition in Fig. 6 due to the much smaller average domain size [37]. Furthermore, we have to point out that a complete description of the evolution of the magnetic properties has to include the influence of the internal stress and the temperature dependence of the magnetostrictive and magnetoelastic energies. The different magnetic behaviors, especially the shape of $M(H)$ curves in Fig. 8, of $Fe_3O_4$ thin films grown on MgO and MAO have been also reported by Bollero *et al.* [44], who considered that the APBs play a very important role in the demagnetization process, that the APBs act as pinning centers for the magnetic domain walls, and that the choice of type of substrate will have a strong effect on the density of APBs [14–17,19–37].

It has been reported that the $Fe_3O_4$ thin films grown on MgO have a large density of APBs [14–16,24] whereas the films on MAO are expected to have a lower density of APBs due to the same structure and closer values of the lattice parameters of the film and substrate. Luysberg *et al.* [32] found that the formation mechanism of APBs in $Fe_3O_4$/MAO is fundamentally different from that in the $Fe_3O_4$/MgO system, and that the formation of misfit dislocations with partial Burgers vectors is responsible for the nucleation of APBs in $Fe_3O_4$/MAO. Hirth *et al.* [66–68] reported that the forces considered to be acting on the dislocations are the elastic interaction force between misfit dislocations, the homogeneous force arising from the difference in atomic volume of the constituents, and the osmotic force produced by the net vacancy flux accompanying interdiffusion. The larger the lattice mismatch between the film and the substrate, the smaller the critical thickness of the film and the higher the density of the misfit dislocations [67,68], which would generate the higher density of APBs [32]. Therefore, it can be suggested that the film on STO (mismatch −7.5%) should have more APBs than the film on MAO (−3.7%). Moreover, the XRD results in Fig. 4 give quite small average domain sizes of 11 nm and 20 nm for 200-nm-thick $Fe_3O_4$ films on STO and MAO, respectively, meaning much larger domain boundaries as compared to those of the film on MgO. It is clear from Figs. 8(a), 8(b) and 8(c) that the film on MgO exhibits incoherent reversal of magnetization while the films on MAO or STO show the rectangular-shaped $M(H)$ curves, which implies that the APBs and the domain size take a small role in the demagnetization process.

It is obtained from Fig. 5 that the 200-nm-thick $Fe_3O_4$ films grown on MgO (001) and MAO (001) show fully strained and fully relaxed states, respectively; the same thickness of film grown on a larger mismatch STO (001) should appear as fully relaxed also [64]. Quite different behaviors of the $M(H)$ curves for $Fe_3O_4$ thin films grown on MgO, MAO, and STO remind us to consider the very important role of strain state in the demagnetization process; the presence of strain actually retards the reversal of domains in applied magnetic field. We therefore only find incoherent change of magnetization around $H_C$ for the film on MgO due to its fully strained state though the average domain size is much bigger than that of the films on MAO and STO. This consideration can be further confirmed by studying the magnetic properties in different thicknesses of $Fe_3O_4$ thin films grown on MAO. Figures 11(a) and 11(b) show the in-plane $M(H)$ curves of 40-nm- and 80-nm-thick $Fe_3O_4$ thin films grown on MAO, respectively. It is clear that the 40-nm thin film exhibits the incoherent reversal of domains like that in Fig. 8(a) while the 80-nm one presents the rectangular-shaped $M(H)$ curves similar to those in Figs. 8(b) and 8(c). From the reciprocal space mapping results, the 80-nm-thick film is nearly fully relaxed whereas the 40-nm-thick one is partial strained; therefore, it is reasonable to consider that the strain state significantly affects the demagnetization behavior of $Fe_3O_4$ thin films. As the thinner film has smaller domain size, we thus observe the larger $H_C$ at temperature around and above $T_V$ for the 40-nm thin film [see Figs. 11(c) and 11(d)], and obvious rapid change of $H_C$ can be seen at the Verwey transition. Moreover, we note that the minimum of $H_C$ is seen at 130 K for the films grown on MAO and STO [see Figs. 10(b) and 10(c)] but at 300 K for the film on MgO [Fig. 10(a)]. As the 200-nm-thick films grown on MAO and STO are fully relaxed, partial magnetic properties, to some extent, are closer to that of the single-crystal bulk though the domain size is much smaller. It has been reported that the absolute value of first-order cubic magnetocrystalline constant $K_1$ decreases rapidly with lowering temperature from 200 K and reaches zero at 130 K (isotropic point) [60], which leads to a broadening of the domain walls because the domain wall width $\delta_W$ is proportional to $K_1^{-1/2}$. As a result, the wall pinning becomes less effective resulting in a decrease of the $H_C$ with reducing the temperature from 200 to 130 K [see Figs. 10(b) and 10(c)], similarly to the single-crystal bulk [59], which, however, is not observed in the film grown on MgO [Fig. 10(a)].

Therefore, it can be confirmed that the strain state significantly influences the magnetic properties in $Fe_3O_4$ thin films.

Furthermore, we found that the in-plane $M_r/M_s$ of the film on MgO is smaller than that on MAO or STO [see Figs. 10(d), 10(e) and 10(f)]. It is theoretically calculated that the $M_r/M_s$ is equal to 0.5 for an assembly of uniaxial particles oriented at random [69], and of 0.831 and 0.866 for positive and negative $K_1$, respectively, of the cubic magnetocrystalline anisotropy materials [70,71]. Dionne reported the relationship between the remanence ratio and the stress, the magnetostriction constants ($\lambda_{100}$ and $\lambda_{111}$), and $K_1$ of the cubic crystal structure; $M_r/M_s$ reduces with increasing tensile stress when $K_1 < 0$ and $\lambda_{100} < 0$ [72,73]. For $Fe_3O_4$ thin films, the $K_1$ and $\lambda_{100}$ are negative above $T_V$ [15,60]; thus the smaller in-plane $M_r/M_s$ of the film on MgO is observed. Similar phenomena are also found in $CoFe_2O_4$ or $NiFe_2O_4$ thin films grown on a piezoelectric substrate [74,75]. Moreover, another property of the in-plane $M_r/M_s$ vs $T$ curve for three samples is also seen in Figs. 10(d), 10(e) and 10(f): a greater $M_r/M_s$ for temperatures below $T_V$ than above $T_V$. It is reported that the easy (hard) axis changes from the [111] ([100]) to [100] ([111]) direction with lowering temperature from high temperature to elow $T_V$ for bulk magnetite [53,54,60], whereas the easy axis always remains in the [100] direction in our thin films at all temperature ranges. It could be considered that the easy magnetization direction may slightly deviate from [100] for very thick film (200 nm) at $T > T_V$, which would influence the remanence ratio; thus the higher $M_r/M_s$ at temperatures below $T_V$ can be observed for very thick film.

To more comprehensively understand the magnetic behaviors in $Fe_3O_4$ thin films, we also investigated the magnetic properties of 40-nm- and 80-nm-thick films on MgO (001); the films on MgO are all fully strained (we thus do not need to consider the influence of strain state) and the $H_c$ and $M_r/M_s$ for three thicknesses as a function of temperature are plotted in Figs. 12(a) and 12(b), respectively. As can be seen all the samples exhibit very sharp jump of $H_c$ at their Verwey transitions, respectively. The value of $H_c$ obviously reduces with increasing thickness below or above $T_V$ due to the great increase of average domain size from 33 nm for 40-nm- to 98 nm for 200-nm-thick $Fe_3O_4$ film, while the demagnetization behaviors are nearly the same for three samples because of the same strain state. It is observed that the change of $M_r/M_s$ at $T_V$ is different for three samples. An obvious increase and decrease of $M_r/M_s$ is seen for 40-nm- and 200-nm-thick films, respectively, whereas the 80-nm-thick one has a special behavior, which presents the decrease of $M_r/M_s$ below $T_V$, like that of the 200-nm film, but a sharp increase of $M_r/M_s$ at $T_V$, similar to that of the 40-nm one, and then remains nearly constant above $T_V$. The absolute difference of $M_r/M_s$ below and above $T_V$ is much smaller than that of 40-nm or 200-nm-thick films. Therefore, it is very interesting that we actually find a clear reversal of $M_r/M_s$ from increase to decrease jump at $T_V$ with increasing thickness, and the critical thickness is about 80 nm. At high temperature, the easy axis is towards the [111] direction for bulk magnetite [53,54,60] but stays in the [100] direction in our thin films. For the thinner $Fe_3O_4$ films, the interface effect is stronger; the easy axis of the domains can be more uniformly oriented to [100], and thus the larger $M_r/M_s$ for the thinner films above $T_V$ can be seen. On the other hand, the interface effect would play a more important role for the smaller anisotropy constant condition, which might lead to the increase in $M_r/M_s$ with rising temperature from below to above $T_V$ for very thin film.

Finally, we note that the $\rho(T)$ curve of $Fe_3O_4$ thin film grown on MgO at the Verwey transition shows a big hysteresis (~4 K) (see Fig. 6); it is expected that the magnetic properties should rapidly change at $T_{V-}$ and $T_{V+}$, and thus a big hysteretic loop can be seen. To observe this phenomenon, we measured $M(H)$ curves with temperature along the cooling down and warming up branches near the $T_V$. The $H_c$ and $M_r/M_s$ as a function of temperature in the vicinity of $T_V$ are plotted in Figs. 13(a) and 13(b), respectively. A big hysteretic loop (~4 K) of the $H_c(T)$ curve with a sharp jump of $H_c$ at $T_{V-}$ and $T_{V+}$, respectively, is clearly observed [Fig. 13(a)]. At temperatures deviating from $T_V$, the $H_c$ is nearly the same for the two different measurement processes. The $M_r/M_s$ vs $T$ in Fig. 13(b) also shows a rapid change at $T_{V-}$ and $T_{V+}$, respectively, despite that the hysteretic loop is not so obvious as the $H_c$. For the Fe$_3$O$_4$ thin films grown on MAO or STO, the hysteretic loop is very narrow (~1 K) and the transition is very broad (Fig. 6); accordingly, the variation of magnetic properties becomes very slow, and it is not easy to distinguish the change of magnetic properties at $T_{V-}$ and $T_{V+}$, respectively.

## VI. CONCLUSION

To summarize, we have systematically investigated the evolution of magnetic properties with temperature in high quality epitaxial Fe$_3$O$_4$ thin films grown on MgO (001), MAO (001), and STO (001) substrates. We observed rapid variations of magnetization, $H_c$, and $M_r/M_s$ at $T_V$, which are consistent with the behaviors of $\rho(T)$ curves for the thin films on different substrates. In particular, we found quite different magnetic behaviors with temperature for the thin films on MgO from those on MAO and STO, in which the domain size and the strain state play very important roles. The $H_c$ is mainly determined by the domain size but the demagnetization process is mainly dependent on the strain state. The small domain size and high strain state result in large coercivity and incoherent reversal of magnetization, respectively. The minimum of $H_c$ at 130 K can be seen for the films on MAO and STO, similar to that of the single-crystal bulk, due to the fully relaxed state, but the smallest $H_c$ is found at 300 K for the films on MgO. Furthermore, we found a smaller remanence ratio for the film grown on MgO than that on MAO and STO, which is induced from the tensile strain effect. Moreover, we observed a very interesting reversal of remanence ratio at $T_V$ with thickness for the Fe$_3$O$_4$ thin films grown on MgO: from a rapid enhancement for 40-nm- to a sharp drop for 200-nm-thick film, and the critical thickness is about 80 nm; the interface effect should play a very important role in this phenomenon. Finally, a clear hysteretic loop of $H_c$ (or $M_r/M_s$) with temperature around $T_V$, corresponding to the loop of the $\rho(T)$ curve, was seen in Fe$_3$O$_4$ thin film grown on MgO. Our work will give a more comprehensive understanding of magnetic behaviors in Fe$_3$O$_4$ thin films.

## ACKNOWLEDGMENTS

The authors would like to thank Prof. Liu Hao Tjeng and Dr. Chun-Fu Chang for useful discussion. The research of X.H.L. was supported by the Max Planck POSTECH Center for Complex Phase Materials.

W.L. and Z.D.Z. were supported by the National Natural Science Foundation of China under Projects No. 51590883 and No. 51331006, and as a project of the Chinese Academy of Sciences with Grant No. KJZD-EWM05-3.

difference in integral profile width between a standard (0.006) and the sample to be analyzed, and $\theta$ is the angle of incidence.

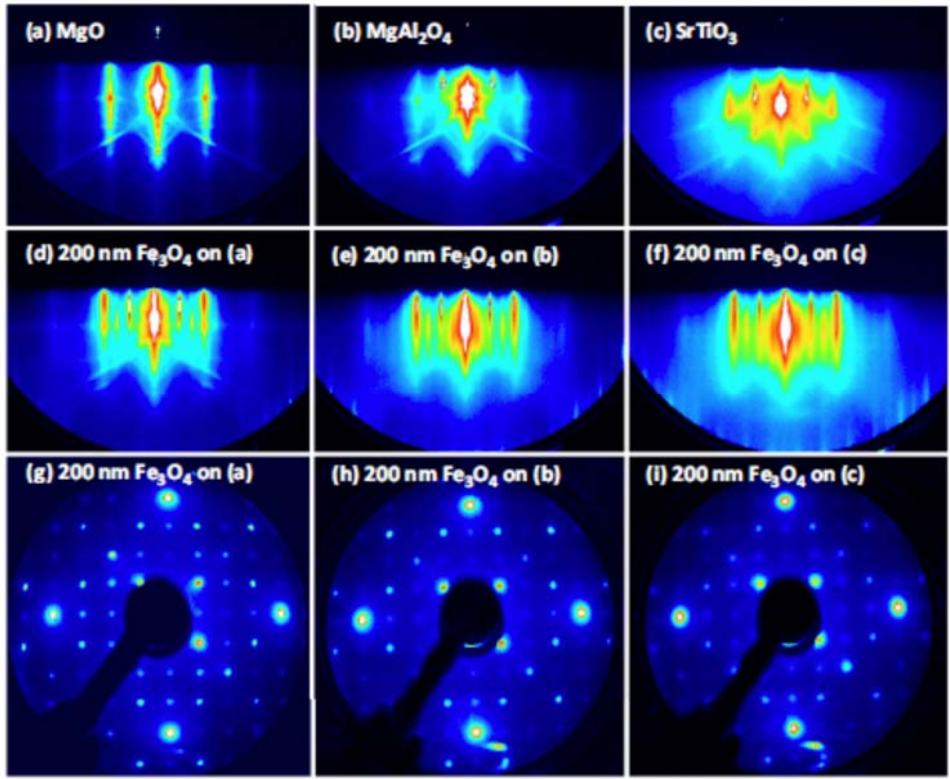

FIG. 1. RHEED electron diffraction patterns of the following: the clean MgO (001) (a), MgAl$_2$O$_4$ (001) (b), and SrTiO$_3$ (001) (c); RHEED and LEED electron diffraction patterns of 200-nm-thick Fe$_3$O$_4$ films grown on MgO (001) (d) and (g), MAO (001) (e) and (h), and STO (001) (f) and (i).

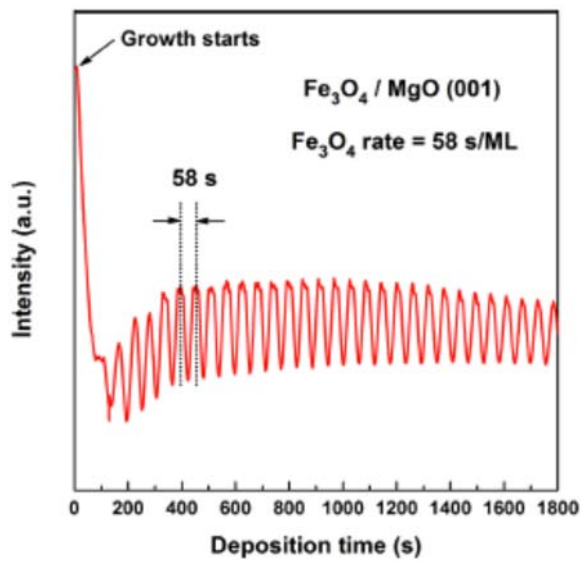

FIG. 2. RHEED intensity oscillations of the specularly reflected electron beam recorded during deposition of Fe$_3$O$_4$ thin film on MgO (001) substrate.

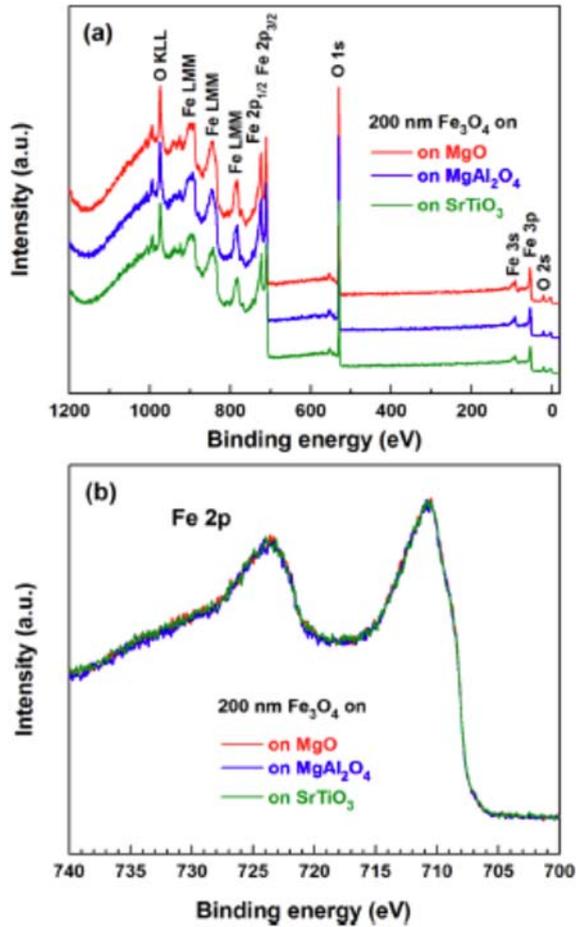

FIG. 3. XPS wide-scan spectra with binding energy from 1200 eV to −18 eV (a); Fe 2$p$ core-level spectra (b) of 200-nm-thick Fe$_3$O$_4$ thin films grown on MgO (001), MgAl$_2$O$_4$ (001), and SrTiO$_3$ (001) substrates.

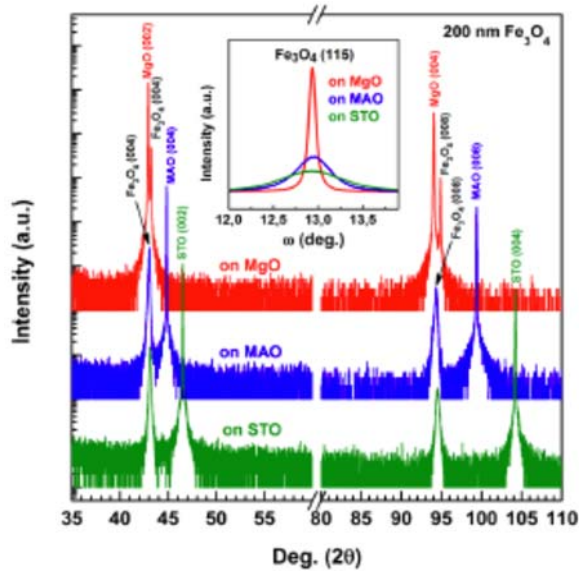

FIG. 4. X-ray diffraction patterns for 200-nm-thick Fe$_3$O$_4$ films grown on MgO (001) (red), MgAl$_2$O$_4$ (001) (blue), and SrTiO$_3$ (001) (olive), respectively. Inset: High-resolution rocking curves of the Fe$_3$O$_4$ (115) peak of these three films.

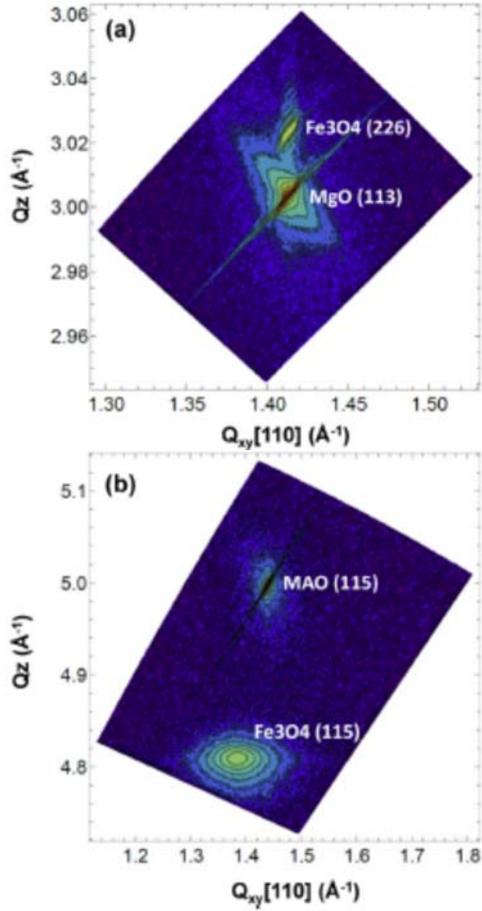

FIG. 5. Reciprocal space mapping in the vicinity of the asymmetrical MgO (113) and Fe$_3$O$_4$ (226) (a), and (115) reflection (b) of 200-nm-thick Fe$_3$O$_4$ films grown on MgO (001) and MgAl$_2$O$_4$ (001), respectively.

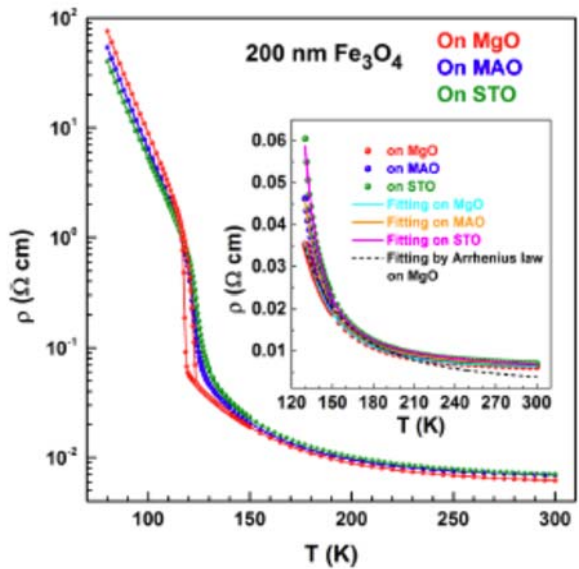

FIG. 6. Resistivity as a function of temperature for 200-nm-thick Fe$_3$O$_4$ films grown on MgO (001), MgAl$_2$O$_4$ (001), and SrTiO$_3$ (001), respectively. Inset: Temperature dependence of resistivity (from 130 to 300 K) and the fitting curves with nonadiabatic polaron model for the three films, as well as the fitting curve by Arrhenius law for the film on MgO (black dashed line).

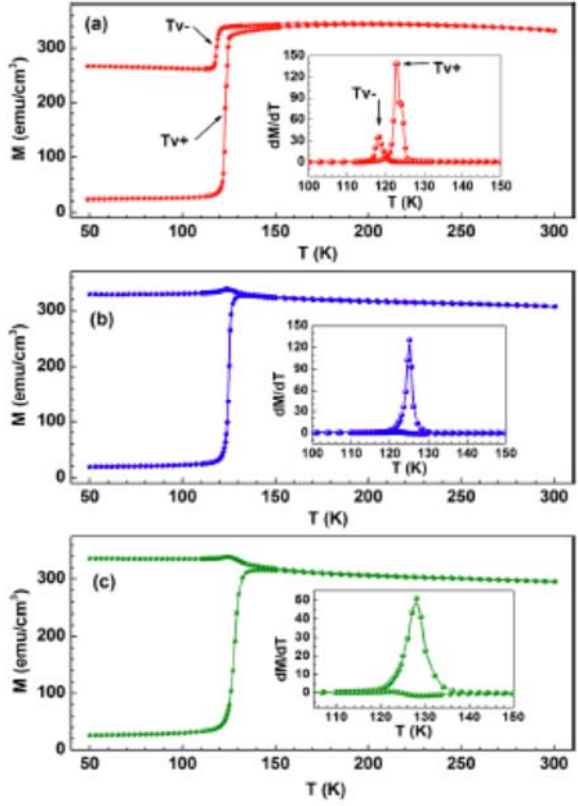

FIG. 7. ZFC-FC temperature-dependent magnetization curve of 200-nm-thick $Fe_3O_4$ films grown on MgO (001) (a), $MgAl_2O_4$ (001) (b), and $SrTiO_3$ (001) (c), in an applied field of 200 Oe. Insets: $dM/dT$ as a function of temperature in the vicinity of Verwey transition temperature.

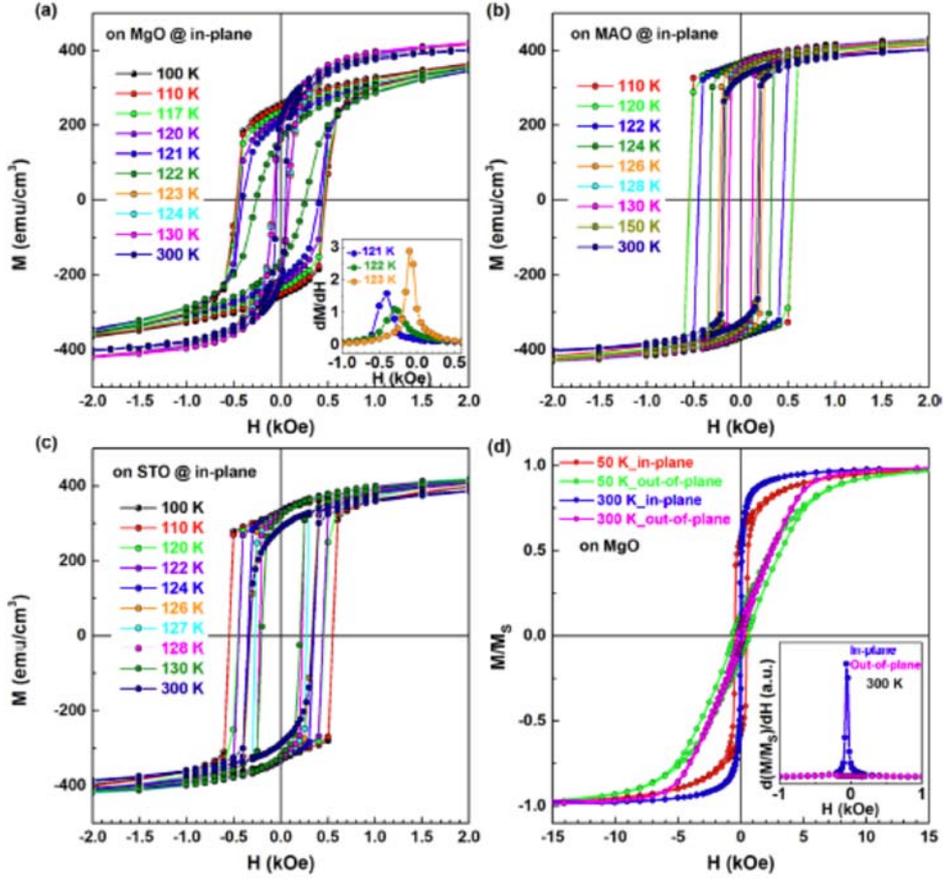

FIG. 8. In-plane magnetic hysteresis loops in applied field of 50 kOe at different temperatures for 200-nm-thick Fe$_3$O$_4$ films grown on MgO (001) (a), MgAl$_2$O$_4$ (001) (b), and SrTiO$_3$ (001) (c), respectively. Inset of (a): $dM/dH$ as a function of magnetic field near the $H_c$ at 121, 122, and 123 K. (d) Normalized magnetic hysteresis loops for in-plane and out-of-plane measured in applied field of 50 kOe at 50 and 300 K, respectively, of 200-nm-thick film grown on MgO (001). Inset: $dM/dH$ dependent on $H$ around the $H_c$ at 300 K for in-plane and out-of-plane conditions.

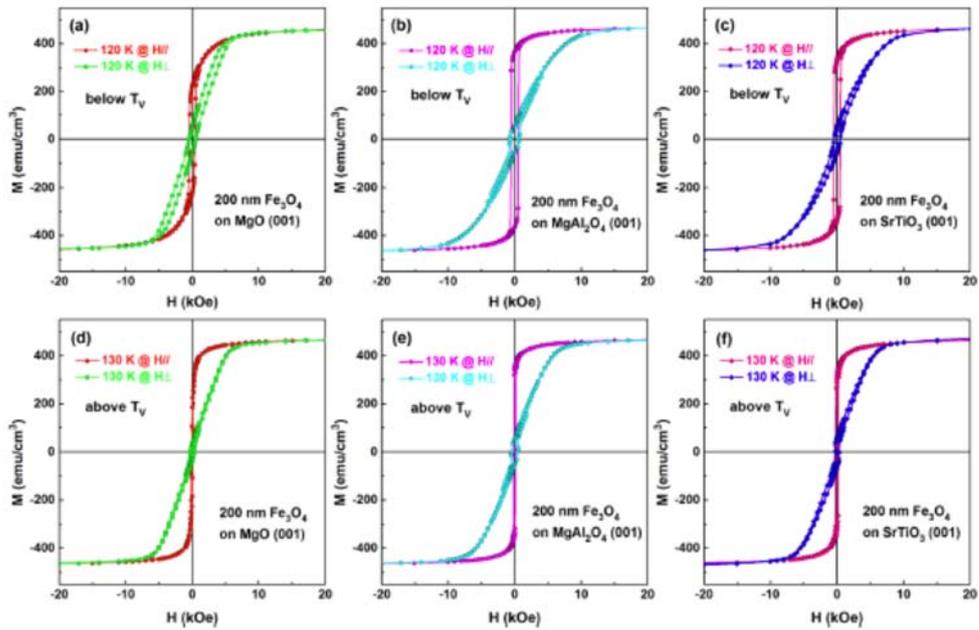

FIG. 9. In-plane ($H_{//}$) and out-of-plane ($H_{\perp}$) magnetic hysteresis loops at 120 K (below $T_V$) and 130 K (above $T_V$) for 200-nm-thick $Fe_3O_4$ films grown on MgO (001) (a) and (d), $MgAl_2O_4$ (001) (b) and (e), and $SrTiO_3$ (001) (c) and (f), respectively.

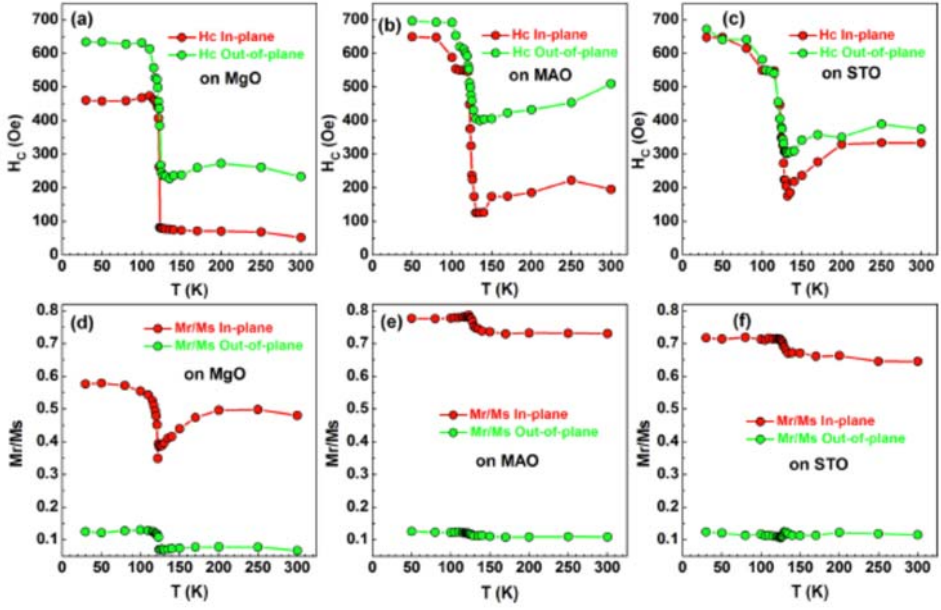

FIG. 10. Evolution of in-plane and out-of-plane $H_C$ and $M_r/M_s$ in dependence on the measuring temperature for 200-nm-thick $Fe_3O_4$ films grown on MgO (001) (a) and (d), $MgAl_2O_4$ (001) (b) and (e), and $SrTiO_3$ (001) (c) and (f), respectively.

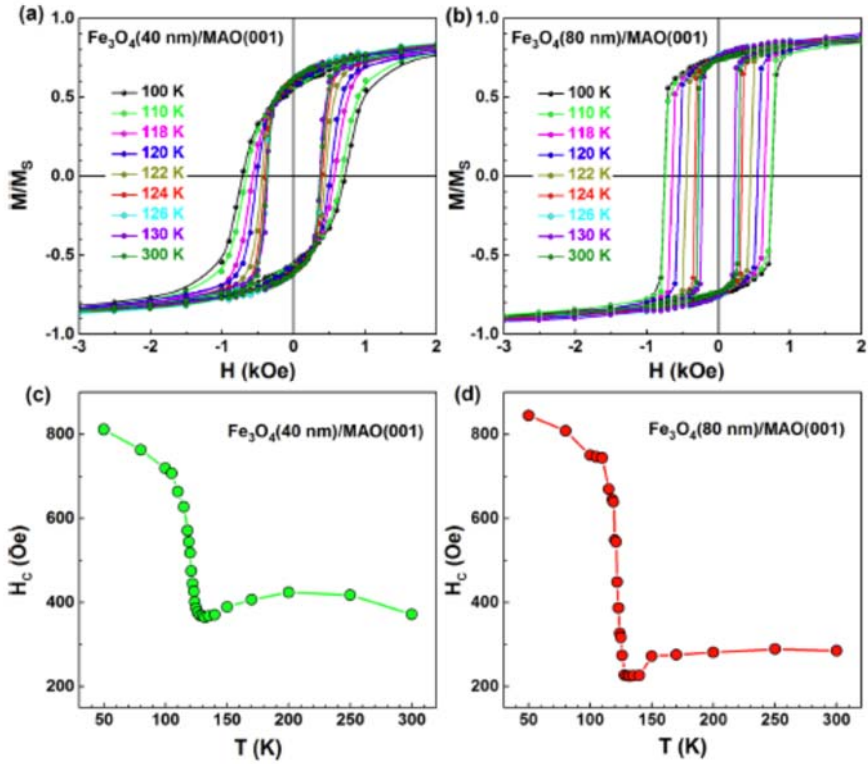

FIG. 11. In-plane magnetic hysteresis loops in applied field of 50 kOe at different temperatures for (a) 40-nm- and (b) 80-nm-thick $Fe_3O_4$ thin films grown on $MgAl_2O_4$ (001). Temperature dependence of in-plane $H_C$ for (c) 40-nm-

and (d) 80-nm-thick $Fe_3O_4$ thin films grown on $MgAl_2O_4$ (001).

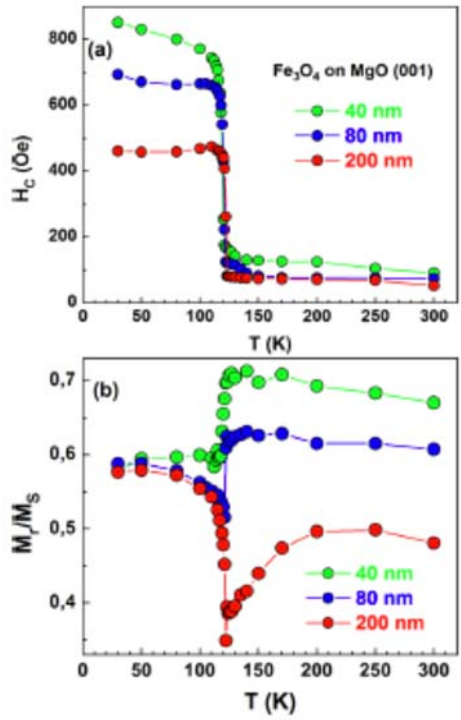

FIG. 12. In-plane $H_c$ (a) and $M_r/M_s$ (b) as a function of temperature for 40-nm-, 80-nm-, and 200-nm-thick $Fe_3O_4$ thin films grown on MgO (001).

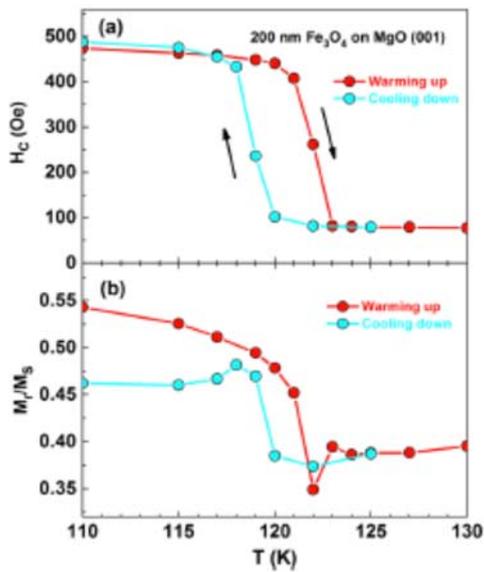

FIG. 13. $H_c$ (a) and $M_r/M_s$ (b) as a function of temperature for cooling down and warming up branches in the vicinity of $T_V$ of 200-nm-thick $Fe_3O_4$ film grown on MgO (001).